\documentclass[useAMS,usenatbib,usegraphicx]{mn2e}

\usepackage{times}
\usepackage{epsfig}
\usepackage{epstopdf}

\def\kms{\relax \ifmmode {\,\rm km\,s}^{-1}\else \,km\,s$^{-1}$\fi}
\def\ha{\relax \ifmmode {\rm H}\alpha\else H$\alpha$\fi}
\def\hb{\relax \ifmmode {\rm H}\beta\else H$\beta$\fi}
\def\hi{\relax \ifmmode {\rm H\,{\sc i}}\else H\,{\sc i}\fi}
\def\hii{\relax \ifmmode {\rm H\,{\sc ii}}\else H\,{\sc ii}\fi}
\def\h2{\relax \ifmmode {\rm H}_2\else H$_2$\fi}
\def\degr{\hbox{$^\circ$}}
\def\arcmin{\hbox{$^\prime$}}

\newcommand{\pab}{Pa$\beta$}

\newcommand{\pii}{[P\,\textsc{ii}]}

\newcommand{\feii}{[Fe\,\textsc{ii}]}

\newcommand{\Hband}{\emph{H}}
\newcommand{\Kband}{\emph{K}}
\newcommand{\etal}{{et al.}}
\newcommand{\cirpass}{\textsc{cirpass}}

\newcommand{\msol}{\mbox{\,M$_\odot$}}        
\def\hmsd#1h#2m#3.#4s{                  
                      \relax
                      \ifmmode #1^{\rm h}\,#2^{\rm m}\,#3\fs#4
                      \else \hbox{$#1^{\rm h}\,#2^{\rm m}\,#3\fs#4$}
                      \fi
                     }
\def\dmsd#1d#2m#3.#4s{                  
                      \relax
                      #1\degr\,#2\arcmin\,#3\farcs#4
                     }

\title[The central region of M83]{The central region of M83: Massive star formation, kinematics, and the location and origin of the nucleus}

   \author[J. H. Knapen et al.]{
   	   J. H. Knapen$^{1,2}$,
	   R. G. Sharp$^{3}$,
	   S. D. Ryder$^{3}$,
           J. Falc\'on-Barroso$^{1,2,4}$,
           K. Fathi$^{5,6}$, 
	   L. Guti\'errez$^{1,7}$\\
$^{1}$Instituto de Astrof\'\i sica de Canarias, E-38200 La Laguna, Tenerife, Spain\\
$^{2}$ Departamento de Astrof\'\i sica, Universidad de La Laguna, E-38205 La
Laguna, Tenerife, Spain\\
$^{3}$Anglo-Australian Observatory, PO Box 296, Epping, NSW 1710, Australia\\
$^{4}$European Space Research \& Technology Centre, Keplerlaan 1, 
Postbus 299, 2200 AG Noordwijk, The Netherlands\\
$^{5}$Stockholm Observatory, Department of Astronomy, Stockholm University, AlbaNova, SE-106 91 Stockholm, Sweden\\
$^{6}$Oskar Klein Centre for Cosmoparticle Physics, Stockholm University,  
SE-106 91 Stockholm, Sweden\\
$^{7}$Universidad Nacional Aut\'onoma de M\'exico, Apartado
Postal 877, Ensenada, B.C. 22800, M\'exico}

\begin{document}

\date{Accepted 2010 June 10}

\pagerange{\pageref{firstpage}--\pageref{lastpage}} \pubyear{2002}

\maketitle

\label{firstpage}

\begin{abstract}

We report new near-IR integral field spectroscopy of the central
starburst region of the barred spiral galaxy M83 obtained with CIRPASS on Gemini-S, which we analyse in conjunction with G\ha FaS Fabry-Perot data, an AAT IRIS2 $K_{\rm s}$-band image, and near- and mid-IR imaging from the {\it Hubble} and {\it Spitzer} space telescopes. The bulk of the current star formation activity is hidden from optical view by dust extinction, but is seen  in the near- and mid-IR to the north of the nucleus. This region is being fed by inflow of gas through the bar of M83, traced by the prominent dust lane entering into the circumnuclear region from the north. An analysis of stellar ages confirms that the youngest stars are indeed in the northwest. A gradual age gradient, with older stars further to the south, characterises the well-known star-forming arc in the central region of M83, and is consistent with a stable scenario where gas inflow into the central regions is facilitated by the galactic bar. 

Detailed analyses of the
\pab\ ionised gas kinematics and near-IR imaging confirm that the kinematic centre
coincides with the photometric centre of M83, and that these are offset significantly, by about 3\,arcsec or 60\,pc, from the visible nucleus of the galaxy. We discuss two possible options, the first of which postulates that the kinematic and photometric centre traces a galaxy nucleus hidden by a substantial amount of dust extinction, in the range $A_V=3-10$\,mag.  By combining this information with kinematic results and using arguments from the literature, we conclude that such a scenario is, however, unlikely, as is the existence of other ``hidden'' nuclei in M83.

We thus concur with recent authors and favour a second option, in which the nucleus of the galaxy is offset from its kinematic and photometric centre. This is presumably a result of some past interaction, possibly related to the event which lies at the origin of the disturbance of the outer disk of the galaxy. We find some indications for a disturbance in the \ha\ velocity field which would confirm the influence of the $m=1$ perturbation in the gravitational potential, but note that further high-quality stellar kinematic data are needed to confirm this scenario.

\end{abstract}



\section{Introduction}

Starbursts provide a significant fraction of the massive star formation (SF) in the
local Universe (e.g., Heckman 1998; Kennicutt et al. 2005), and most probably an even higher fraction at earlier epochs (e.g., Heckman 2005 for a review). The
combined energy output from their massive stars can cause superwinds, which can
enrich the interstellar, and even the intergalactic, medium (as reviewed in detail by Veilleux, Cecil, \&
Bland-Hawthorn 2005). Some of the main open questions relating to starburst
activity in galaxies are the combination of physical conditions which can lead
to the triggering of the burst, how long a starburst can be sustained for, and
how a starburst is fuelled. 

Even though the definition of a starburst is not at all straightforward (Knapen \& James 2009), the spiral galaxy M83 (=NGC~5236) is often considered a starburst host. M83 is a nearby galaxy (4.5~Mpc, Thim et al. 2003, which implies that 1\,arcsec corresponds to 22\,pc) of morphological type SAB(s)c (de Vaucouleurs et al. 1991; Buta et al. 2007), and has a prominent bar with equally prominent dust lanes seen to connect the central region, through the bar, to the disk.

The atomic hydrogen in M83 is well known to be extended much further than the optical disk (e.g., Huchtmeier \& Bohnenstengel 1981), with a central depression, a relatively bright inner area corresponding roughly to the extent of the optical disk, and a fainter extended disk, up to a degree in size. The latter shows evidence for streamers, tidal arms, and kinematic disturbances, which indicate a past and/or present interaction, possibly with the close neighbor galaxy NGC~5253 (Park et al. 2001 [with an illustration of their \hi\ integrated intensity map as Fig.~2 of Hibbard et al. 2001]; Miller et al. 2009). Thilker et al. (2005) reported from {\it Galex} UV imaging that stars are forming in some of these outer \hi\ arms, and a newer {\it Galex} image (F. Bigiel et al., in preparation)\footnote{http://www.galex.caltech.edu/newsroom/glx2008-01r.html} shows that SF in fact occurs in almost all the \hi\ arms and filaments, out to very large distances from the main optical disk.

In this paper, we study the central, starbursting, region of M83.  In a
detailed near-IR study of the nuclear region of M83, Gallais et al. (1991)
found a bright red nucleus and an arc of massive SF to the SW, aspects of which had been seen earlier by, e.g., Telesco (1988) and Wolstencroft (1988). 
Elmegreen et al. (1998) reported the presence of a double ring structure, which
they identified with the inner and outer inner Lindblad resonances (ILRs).
Observations of the bright CO emission allowed Sakamoto et al. (2004) to interpret
the inner ring as a rotating gas disk (see also Lundgren et al. 2004). Fathi et
al. (2008), from new H$\alpha$ Fabry Perot imaging of the galaxy, 
detail a nuclear disk which is rotating faster than the outer disk, and find gas 
which is spiralling into the nuclear region from the bar, fuelling the starburst. 

On the basis of near-IR spectroscopy of the \Hband\ and \Kband\ CO
absorption bands and the Br\,$\gamma$ emission line, Puxley et
al. (1997) proposed an age gradient in the SW starburst ring.  This
was confirmed later with \emph{Hubble Space Telescope} ({\it HST}) WFPC2 imaging by
Harris et al. (2001), who modelled the SF history for the stellar
clusters in the region, and also spectroscopically by Thatte et al. (2000), Ryder et al. (2005), and Houghton \& Thatte
(2008). An age gradient in the ring is {\it prima facie} evidence for
organised, as opposed to randomly occurring, SF, and thus confirms a
picture of bar-driven inflow (e.g., Allard et al. 2006; B\"oker et
al. 2008; Mazzuca et al. 2008).

Thatte et al. (2000) and more recently Sakamoto et al. (2004), Mast et al. (2006),
D\'\i az et al. (2006), and Rodrigues et al. (2009) studied the velocity structure in the
circumnuclear region and interpret it as evidence for a second,
dust-extinguished, nucleus of M83.
Thatte et al. (2000) refer to the double nucleus of M83, propose that this second nucleus is the ``true'' one, and
suggest the ``optical nucleus'' may be the trigger for the starburst
activity in M83. Houghton \& Thatte (2008), however, cannot confirm the local maximum in the stellar velocity dispersion at the location of the ``second nucleus'' on which Thatte et al. (2000) based their conclusions. Instead, Houghton \& Thatte (2008) conclude that there is no evidence for the existence of this ``second nucleus'', nor, in fact, for any other additional nuclei, and that the offset between the positions of the visible nucleus and the photometric and kinematic centre of M83 is likely due to an $m=1$ perturbation in the gravitational potential, in turn possibly related to a past interaction. 

In the present paper we concentrate on the properties of the massive SF in the central region of M83, and on the location and origin of the nucleus. We do this primarily on the basis of a new data set of near-IR integral field spectroscopy, obtained using the CIRPASS instrument on the Gemini South telescope, but using also a new $K_{\rm s}$ image from the Anglo-Australian Telescope (AAT) and archival data from the {\it HST} and from the {\it Spitzer Space Telescope} ({\it SST}). We describe the data and their reduction in Sect.~2, and the overall properties of the circumnuclear region in Sect.~3. The nucleus of M83 is examined in Sect.~4, and in Sect.~5 we discuss critically where the true nucleus of M83 is.  Sect.~6 then summarises our results.


\section{Observations and data reduction}


\subsection{CIRPASS spectroscopy}

\begin{figure*}
\begin{center}
\includegraphics[width=0.95\textwidth]{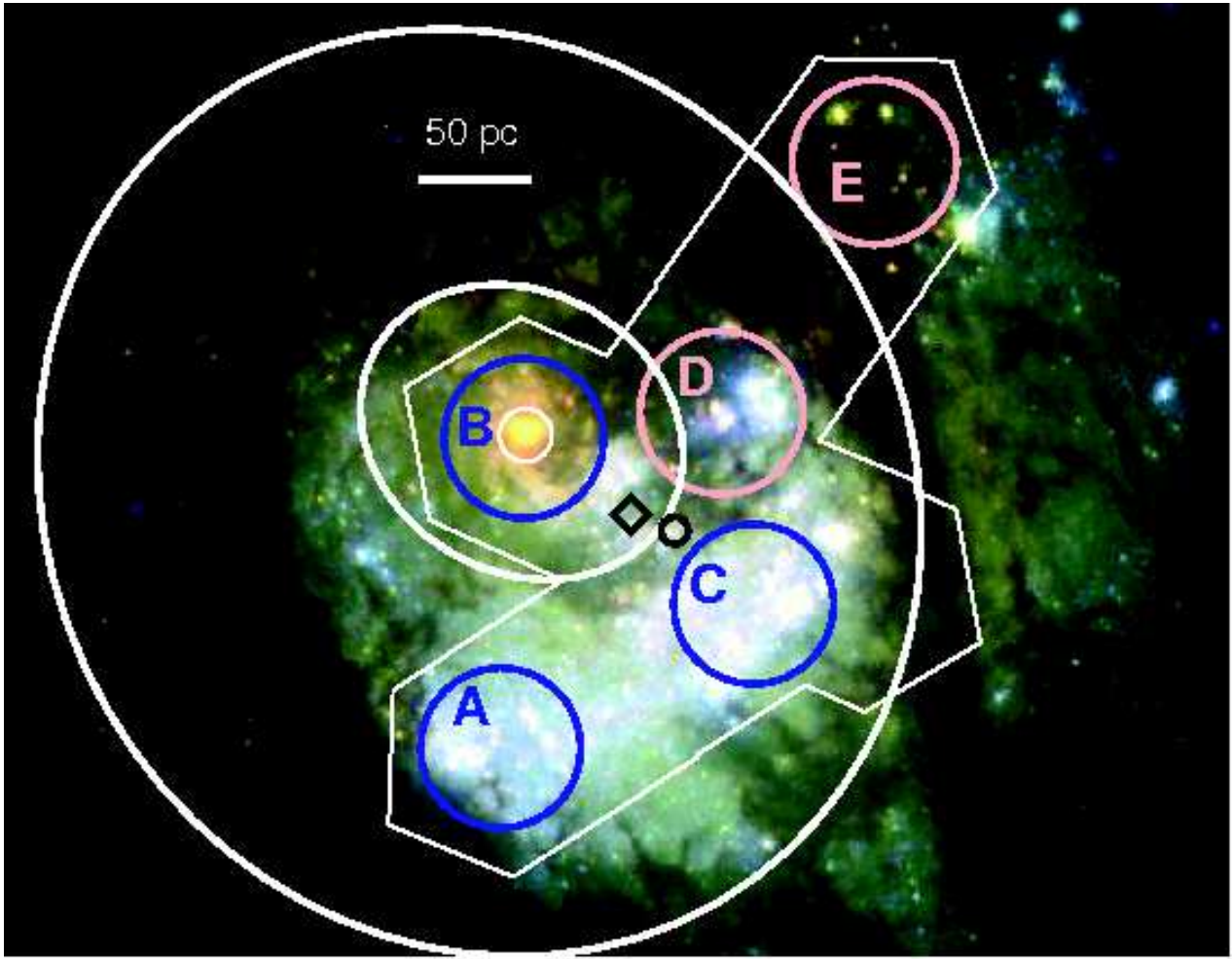}
\end{center}
  \caption{The {\it HST} true-colour image (F300W,
  F547M and F814W filters) from Harris et al. (2001) with the outline of
  our \cirpass\ observations overlaid. North is up, East to the left.
  The ``optical nucleus" is marked with a white circle, the \pab\ kinematic and photometric centre of the galaxy with a black diamond and circle, respectively, and the labelled
  regions indicate the apertures from Ryder et al. (2004). The white ellipses indicate the inner and outer dust rings identified by Elmegreen et al. (1998). The scale indicated, of 50\,pc, corresponds to 2.3\,arcsec on the sky. The diameter of the apertures is 3.2\,arcsec.}
 \label{cirpass_field}
\end{figure*}

\begin{figure*}
\begin{center}
\includegraphics[width=0.45\textwidth]{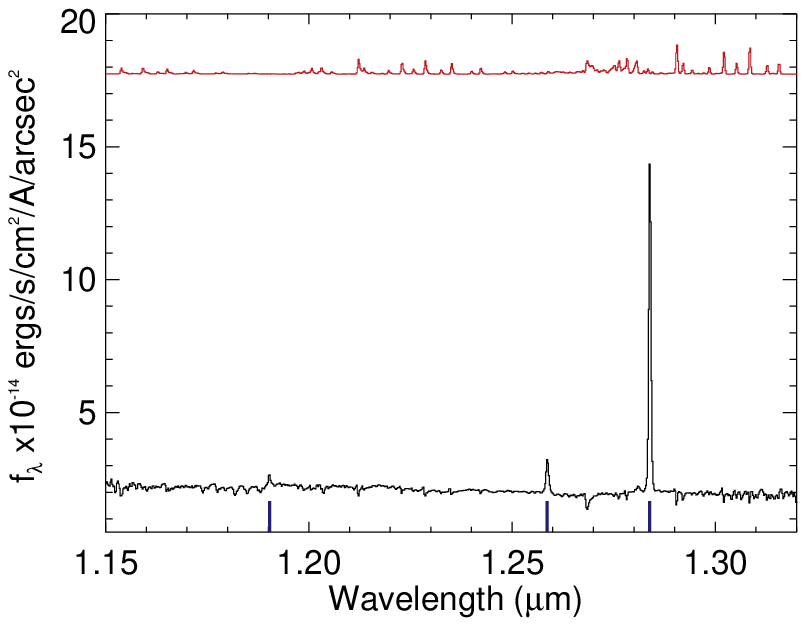}
\includegraphics[width=0.45\textwidth]{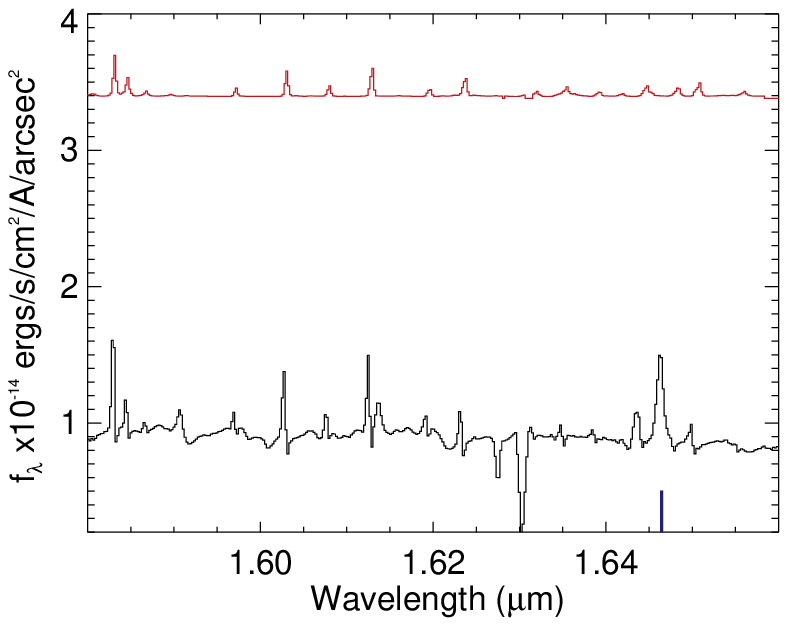}
\end{center}
  \caption{Sample spectra, corresponding to the emission integrated over aperture D (as defined in Fig.~\ref{cirpass_field}), in the $J$ (left) and $H$ (right) bands. Key spectral lines (see Sect.~2.1) are identified by vertical tickmarks. An arbitrarily shifted sky spectrum is shown as the red (upper) spectra.}
 \label{spectra}
\end{figure*}

Near-IR integral field observations of three overlapping regions of
the circumnuclear starburst area (Fig.~\ref{cirpass_field}) of M83 were made with the
Cambridge Infra-Red Panoramic Survey Spectrograph (\cirpass, Parry et
al. 2004) at the Gemini South telescope, during 2003 March 9 and 10
(Ryder et al. 2004; Sharp et al. 2004), under $\sim$1\,arcsec seeing
conditions. We reduced the data using the \cirpass\ IRAF package and
standalone software (see Krajnovi\'c et al. 2007 for a more extensive description).  The spectra were
marginally undersampled at FWHM$=1.8$\,pix on the Hawaii 1K near-IR array
detector. The 2.2\,\AA\,pix$^{-1}$ dispersion yields a resolution of
$\sim$4\,\AA, or $R\sim3000$ in the $J$-band
($\lambda=1.138-1.348\,\mu$m) and $R\sim4000$, in the $H$-band (1.450-1.668\,$\mu$m, where the red limit was set by a 1.67\,$\mu$m blocking filter to reduce the thermal background), at 0.36\,arcsec per hexagonal lens
element (re-sampled during the data reduction to 0.20\,arcsec). The $J$-band spectra include the \pab\ (1.2818\,$\mu$m), \feii\ (1.2567\,$\mu$m), and \pii\ (1.1882\,$\mu$m)
emission lines (rest wavelengths from Oliva et al. 1990, in M83 these lines are redshifted by 0.0022\,$\mu$m assuming a recession velocity of 513\,\kms), as well as the weak 1.279\,$\mu$m He{\sc i} line (3D-5F), at practically identical wavelength to the 1.2788\,$\mu$m \feii\ line (6D3/2 - 4D3/2), on the blue shoulder of the \pab\ emission. The $H$-band spectra include the emission line of \feii\ at 1.6435\,$\mu$m and the CO (6, 3) absorption band-head at 1.6\,$\mu$m. The emission line of Pa\,$\gamma$ fell outside our observed wavelength range.
All emission lines are well approximated as single Gaussian emission
features at the limit of the instrumental resolution, although the limited signal-to-noise ratios outside a few individual bright regions restricted our analysis to the \pab\ line. Sample spectra, corresponding to the emission observed from aperture D in the $J$- and $H$-bands, are presented in Fig.~\ref{spectra}. Uncertainties were estimated based on a bootstrap re-sampling of the spectra from each aperture, making the assumption that the underlying measurement error dominates over the goodness of fit to the final spectral line in a single spectrum. The errors indicated in Table\ref{ages} 
are 1-sigma values from the bootstrap distribution, and look roughly Gaussian in each case.


\subsection{$K_{\rm s}$-band image}

We obtained a wide-field $K_{\rm s}$-band image of M83 with the IRIS2
instrument (Tinney et al. 2004) on the AAT
during the night of 2003 June~16 UT in 2.2~arcsec seeing. Nine
dithered exposures of 56~sec were interleaved with similar exposures
on blank sky 10~arcmin away. Data reduction was carried out using
the ORAC-DR pipeline (Cavanagh et al. 2003) within the {\sc starlink}
package using the CHOP$\_$SKY$\_$JITTER recipe. Preprocessing of all raw frames
included the subtraction of a matching dark frame, linearity and 
interquadrant crosstalk correction, and bad pixel masking.

\begin{figure}
\includegraphics[width=0.5\textwidth]{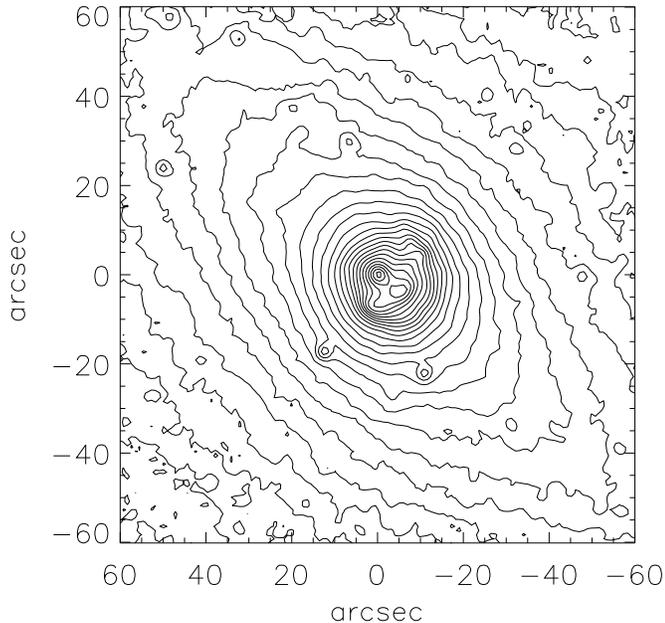}
\caption{Central area of the $K_{\rm s}$-band image of M83. Contours are separated by 0.2\,mag\,arcsec$^{-2}$, and span a range from 17 to 13\,mag\,arcsec$^{-2}$. N is up, E to the left.}
\label{kmap}
\end{figure}

The modal pixel value of each sky frame was calculated, then all sky 
frames were offset in intensity to a common modal value (an additive 
offset rather than a simple scaling was used because the frames have 
yet to be flatfielded at this stage, so a change in sky intensity between frames will 
not result in the counts for every pixel going up or down by the same 
factor). A flatfield image was formed from the median value at each 
pixel of the sky frames, then all sky and object frames were 
flatfielded. Next the modal pixel values of each of the two sky frames 
bracketing each object frame were averaged, and subtracted from that 
object frame. The nine object frames were aligned
using field stars, and mosaiced to yield the image shown in
Fig.~\ref{kmap}. Photometry of 2MASS stars in the field of view has been
used to provide a photometric calibration.


\subsection{{\it SST} IRAC imaging}

\begin{figure*}
\includegraphics[width=0.9\textwidth]{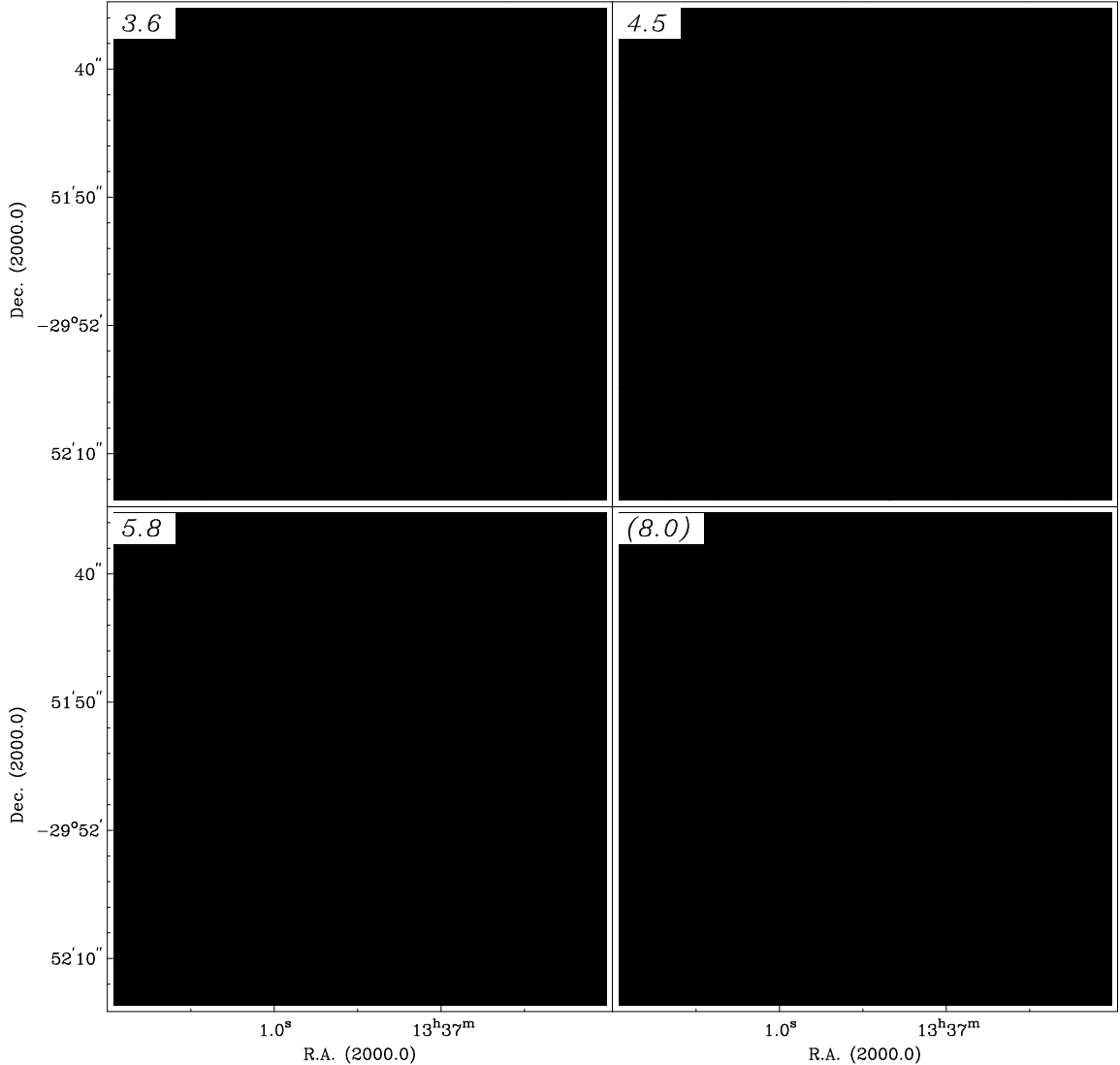}
\caption{{\it SST} IRAC 3.6, 4.5, and 5.8\,$\mu$m images of the central
region of M83, as well as ({\it bottom right}) the dust-only 8\,$\mu$m
image (see Sect.~2.3). Contours are plotted logarithmically
at 4.0, 6.3, 10, 15.8, 25.1, 39.8, 63.1, and 100\,MJy\,sr$^{-1}$
for the 3.6\,$\mu$m image, at 2.5, 4.0, 6.3, 10, 15.8, 25.1, 39.8, and 63.1\,MJy\,sr$^{-1}$
for the 4.5\,$\mu$m image, and at 6.3, 10, 15.8, 25.1, 39.8, 63.1, 100, 158.5, and 251.3\,MJy\,sr$^{-1}$
for the 5.8\,$\mu$m image. The
visible nucleus of the galaxy is indicated with a cross.
}
\label{spitzer}
\end{figure*}

\begin{figure*}
\includegraphics[width=0.9\textwidth]{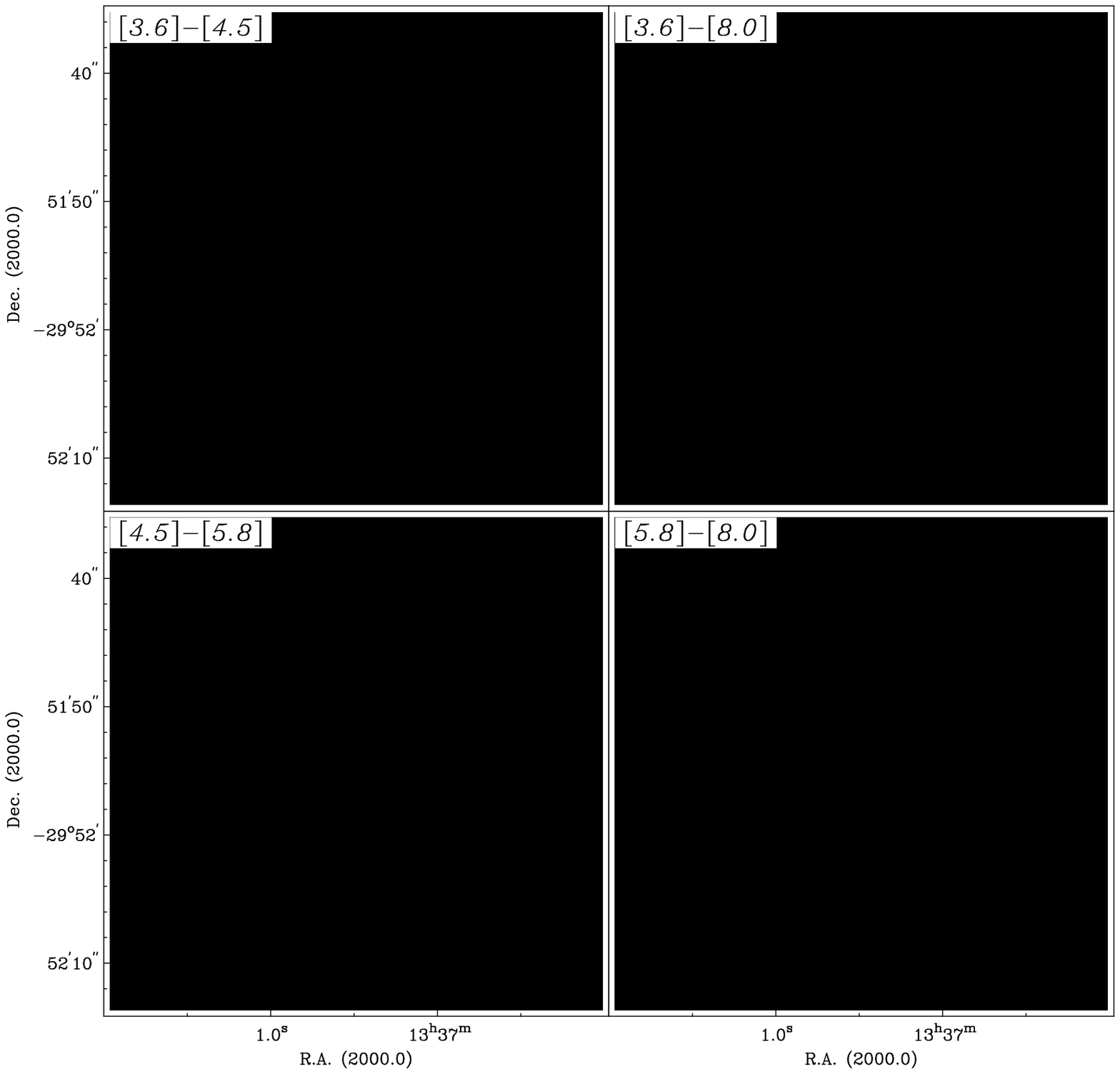}
\caption{{\it SST} IRAC colour index  maps of the central region of
  M83. In all maps, blue tones
indicate blue colours (e.g., more 3.6 than 4.5\,$\mu$m emission), while
on the other extreme of the range, red and white tones indicate the
reddest colours. The visible nucleus of the galaxy is indicated with a
cross.
}
\label{spitzer_ratios}
\end{figure*}

We retrieved {\it SST} (Werner et al. 2004) Infrared Array Camera (IRAC; Fazio et al. 2004) imaging data of M83 from the
archive, data which were obtained as part of G. Rieke's GTO programme, ID 59. We rotated the images
to a N-E orientation to be compared directly with our
other data, and produced colour index images. The latter
were made in magnitudes, by taking the logarithm of the ratio of two
images and multiplying that by 2.5. Following the prescription of Pahre et al. (2004) and Calzetti et al. (2005), we produced a ``dust emission" image at 8\,$\mu$m by subtracting the 3.6\,$\mu$m stellar image. The pixel scale is 1.2~arcsec. The spatial
resolution is just over 2 pixels in each band.  We present IRAC images
in the 3.6, 4.5, 5.8 and dust-only 8.0\,$\mu$m bands of the central region of M83
in Fig.~\ref{spitzer}, and colour index images in Fig.~\ref{spitzer_ratios}.


\subsection{{\it HST} imaging}

We obtained {\it HST} near-IR imaging taken with the NIC2 camera of NICMOS from the archive. We used an image taken through the F222M filter, at a central wavelength of 22181\,\AA, which was obtained as part of programme number 7218, by M.~J. Rieke and her team. The image was taken on 1998 May 16, and has an exposure time of 176\,s. The pixel scale is $\sim0.075\times0.075$\,arcsec$^2$. 

We used individual images as calibrated by the CALNICA pipeline, version 4.4.0, in the {\it HST} archive. We manually aligned and combined these, and subtracted a residual background contribution (of 0.7\,DN) to the emission estimated by scaling the outer part of the surface brightness profile of the NICMOS image (between 8 and 11\,arcsec in radius) to that derived from the AAT image.  For photometric calibration, we used the information given in the header of the NICMOS image.


\section{Structure of the central region}

\begin{figure}
\includegraphics[width=0.45\textwidth]{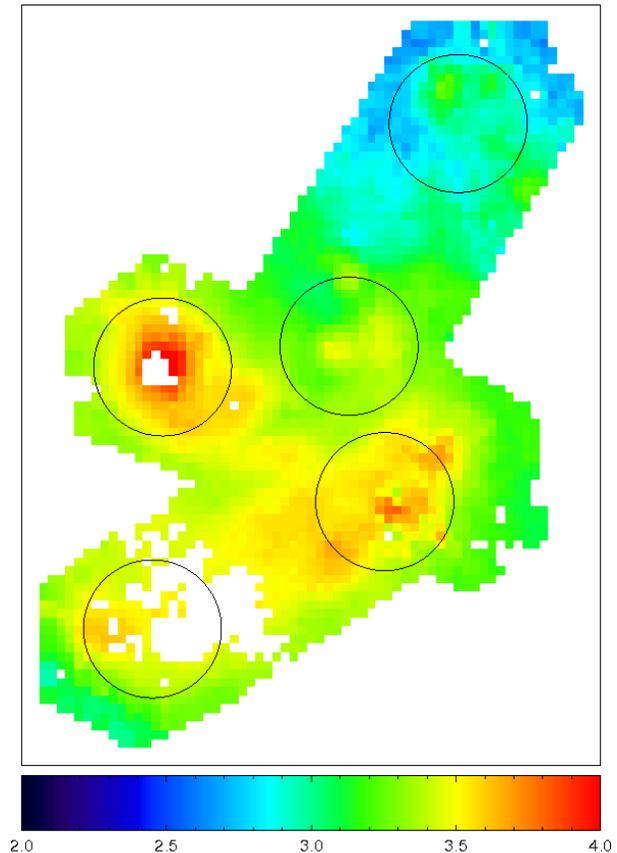}
  \caption{Pa$\beta$ continuum map. The colour bar indicates values as the log of the integrated intensity, where intensity is in units of $10^{-17}$\,ergs/s/cm$^2$/\AA/(CIRPASS lens area), with the CIRPASS lenses measuring 0.2\,arcsec squared. The five circles indicate apertures A-E as identified in Fig.~\ref{cirpass_field}.}
\label{PaB_cont}
\end{figure}

\begin{figure*}
\includegraphics[width=0.95\textwidth]{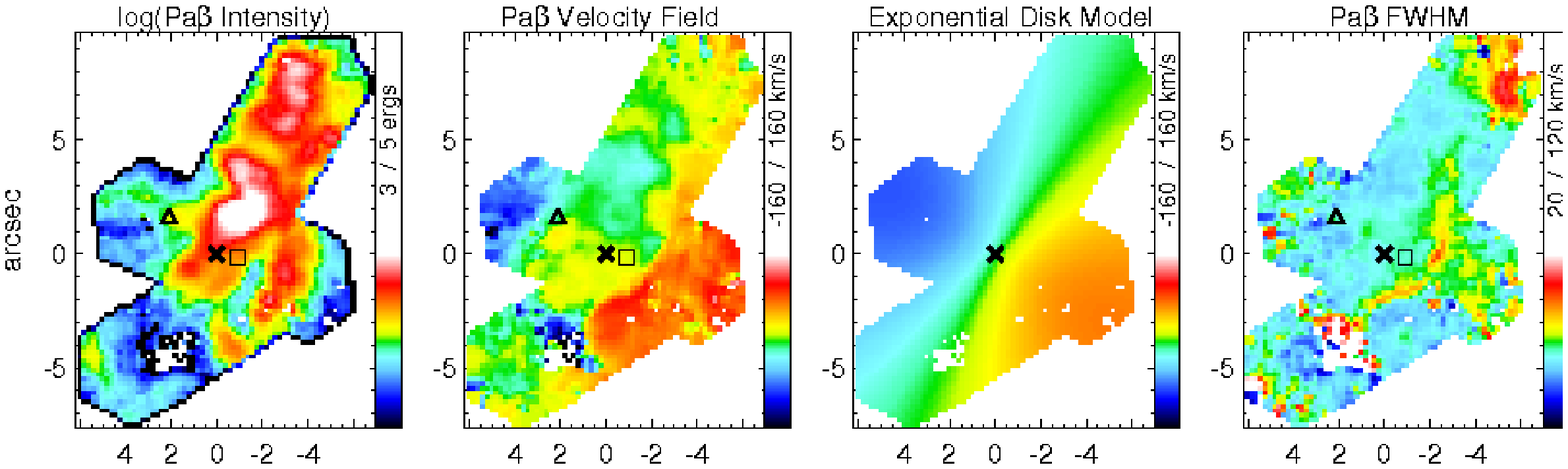}
  \caption{From left to right: \pab\ total intensity map, velocity field, reconstructed inner disk model from Fathi et al. (2008), and velocity dispersion map of the central region of M83 derived
from our \cirpass\ data. Pa$\beta$ intensities are in units of
10$^{-17}$\,erg\,s$^{-1}$\,cm$^{-2}$, and all panels have
the same orientation as Fig.~\ref{cirpass_field}. The {\it triangle} marks the position
of the visible nucleus, the {\it square} that of the kinematic nucleus as
derived by Fathi et al. (2008), and the {\it cross} the rotation centre of the \pab\ velocity
field. }
\label{PaB}
\end{figure*}

The interplay between the enhanced SF and the dust gives rise to
interesting differences between the optical and near-IR continuum, and Pa$\beta$
views of the central region. Dust structure is clearly outlined in the optical
(Fig.~\ref{cirpass_field}), and regions of high dust-extinction (such as E, as identified in
Fig.~\ref{cirpass_field}) become more prominent in the near-IR. This can, for instance, be seen in the continuum image at wavelengths near that of the \pab\ line as derived from our data set (Fig.~\ref{PaB_cont}). The near-IR and, to a somewhat lesser extent, optical continuum images are dominated by the ``visible nucleus''---the region we marked B in Fig.~\ref{cirpass_field}, and the star-forming arc to the W and SW thereof. 

The Pa$\beta$ line emission map (Fig.~\ref{PaB}, left panel) is strikingly different from the continuum view. It
shows how the visible nucleus (B) is currently almost completely devoid of massive star
formation, and how the regions to the N and NW of the centre (D and E) are the most
prominent. Region A (to the SE) is practically absent in Pa$\beta$, in contrast
to the visible (Fig.~\ref{cirpass_field}) where it is as prominent as the main star-forming arc to
the SW (C), which is itself rather inconspicuous in Pa$\beta$. This indicates
that the phase of massive SF may have ended in region A, as we will see confirmed from the age dating analysis in Sect.~3.2. On the basis of our new data, and making reference where relevant to published work in the literature, we will now describe some salient features of the central region, emphasizing the morphology of SF and dust, and the recent history of the SF.


\subsection{Dust structure}

The overall structure of the central region has been well studied. It is dominated in visible light by the star-forming arc, where massive SF can be seen at many wavelengths, including that of \ha. This area of massive SF is seen to extend to the north in the near-IR, e.g., in our \pab\ map (Fig.~\ref{PaB}). In the visible, the SF to the north of the nucleus is mostly obscured by the dust lane which on larger-scale images (e.g., Fig.~1 of Houghton \& Thatte 2008) can be traced from the spiral arms in the disk, through the NE part of the large bar of M83, and into the central region where it cuts through the star-forming arc between our regions D and E (Fig.~\ref{cirpass_field}). An equivalent dust lane can also be seen in the spiral arms to the north of the central region, which curves around and continues through the SW part of the large bar. This dust lane cradles the star-forming arc on the SE side (see Fig.~\ref{cirpass_field}), delimiting the area of massive SF rather than cutting through it. 

The set of dust lanes circling into the central region gives rise to the outer of the pair of circumnuclear dust rings identified by Elmegreen et al. (1998) on the basis of near-IR colour index imaging. These authors also identify an inner dust ring. Both these rings can be recognised, and have been marked, in our Fig.~\ref{cirpass_field}. Whereas we agree with the resonant origin of the outer of these two ring features, we think that the inner dust ring may be shaped as it is by an extended patch of dust, punctured by small areas of a combination of SF and less dense dust. Our main argument for this is that the inner ``ring'' is not only not centred on the visible nucleus, but also not on the photometric and kinematic centre (Fig.~\ref{cirpass_field}).

Elmegreen et al. (1998) also identify a small linear red (interpreted as dust) feature as an inner bar, and postulate that this may pinpoint the location of the path for gas to flow to the ``central starburst''. In contrast, we claim that the origin of this linear feature lies in the same combination of local variations in dust and stellar densities remarked upon in the preceding paragraph. In addition, the ``starburst'' activity of M83 originates in the star-forming arc which lies well outside this putative inner dust bar, and the inflow of gas to this star-forming region is through the main dust lanes in the large bar which dominates the visible/near-IR  view of the central 5-10\,kpc of M83.


\subsection{Star formation: age gradient in the star-forming arc} 

\begin{figure}
\includegraphics[width=0.45\textwidth]{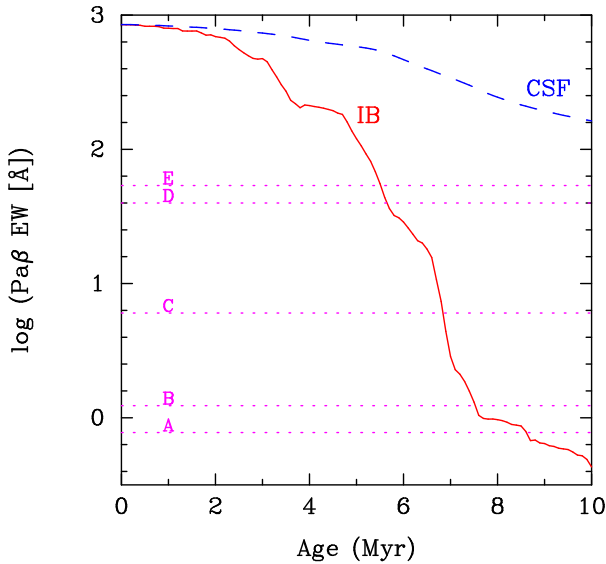}
  \caption{Values for the \pab\ EW for the five apertures as defined in Fig.~\ref{cirpass_field} compared to evolutionary tracks from Leitherer et al. (1999) for the limiting cases of an instantaneous burst (IB, red continuous curve) and continuous SF (CSF, blue dashed curve). Formal uncertainties in the measurements are smaller than the thickness of the lines used to denote them in the Figure, but uncertainties in the models, arising from a variety of assumptions, will be considerably higher.}
\label{ages_pb}
\end{figure}

The existence of an age gradient with the oldest stars in the southern part of the star-forming arc, and the youngest ones to the north has been reported in the literature
(Puxley et al. 1997; Thatte et al. 2000; Harris et
al. 2001; Ryder et al. 2005; Houghton \& Thatte 2008; see the latter paper for a review of earlier work by others which reported an age gradient in the opposite sense to what is found by the more recent authors quoted above). In this paper, we show how the \pab\ equivalent width (EW) allows us to determine fairly unambiguous ages, and indeed confirm the age gradient.

The signal-to-noise ratio in each pixel of our data does not allow us to replicate the analysis for individual clusters, as performed by Harris et al. (2001). Instead, we determined values averaged over the apertures A-E as identified in Fig.~\ref{cirpass_field}. We use the EW for the Pa$\beta$, and list the values in Table~\ref{ages}. 

Figure~\ref{ages_pb} shows the location of the EW(\pab) points in comparison with evolutionary tracks from the Starburst99 model suite (Leitherer et al. 1999). The tracks assume a Salpeter-like initial mass function with exponent $\alpha=2.35$, an upper mass cutoff of $100\,\msol$, and a twice-solar 
metallicity ($Z=0.04$, consistent with the observed radial oxygen 
abundance gradient; Zaritsky et al. 1994).. The {\it first conclusion} we can draw from this figure is that {\it a continuous SF model can be excluded}, although formally one could contemplate a continuous model for apertures D and E (the curve for continuous SF intercepts the lines for aperture E and D at an age of around $10^8$\,yr, that for C at around $10^{10}$\,yr, and those for B and A at even higher values). 

The location of the points allows us to determine ages in the case of an instantaneous burst, and these ages are given in Table\ref{ages}. Given the uncertainties in our \pab\ EWs (of 0.02-0.03 in the log) our ages are, in theory, good to around 0.01\,Myr. But the Starburst99 models use a number of assumptions, and we will adopt conservative error margins of $\pm0.1$\,Myr in our ages. The determined age values are in broad agreement with an age range of $5-10$\,Myr as reported by Harris et al. (2001) and confirmed by Houghton \& Thatte (2008). 

\begin{table}
\begin{center}
\begin{tabular}{ccc}
\hline
\noalign{\smallskip} 
Region & log EW(\pab) & Age\\
         & log(\AA) & Myr\\
\hline
\noalign{\smallskip}
A & $-0.11\pm0.02$ & 8.6\\
B & $0.09\pm0.03$ & 7.5\\
C & $ 0.78\pm0.02$ & 6.8\\
D & $1.60\pm0.02$ & 5.7\\
E & $1.73\pm0.02$ & 5.5\\
\noalign{\smallskip}
\hline
\end{tabular}
\end{center}
   \caption[]{\pab\ EW values and derived ages, in Myr, of the stellar populations at the location of the five regions, A--E. Uncertainties in the ages are estimated to be 0.1\,Myr.}
\label{ages}
\end{table}

A {\it second conclusion} is that we can {\it confirm the previously reported age gradient} in the star-forming arc.
Our analysis yields the lowest ages for region E, the
dust-obscured SF site to the NW.  There is a clear gradual
age gradient from the NW region E, via the zones D, and C, to A, with ages
increasing gradually from 5.5 (E) to 8.6\,Myr (A). The age gradient in itself mostly explains
why the northern part (D and E) of our \cirpass\ field is much brighter in
Pa$\beta$ than the southern part (A and C), while the need for near-IR spectroscopy
is obvious from the presence of the dust in this region. 

The NW star-forming region (E, the youngest, see above) is located where the
northern dust lane spirals in from the bar and disk of the galaxy to the
nuclear starburst region (Fathi et al. 2008). The dust obscures most of the SF in 
regions D and E as seen in the optical (Fig.~\ref{cirpass_field}), whereas these regions, and 
especially D, are the dominant features in the Pa$\beta$ emission map (Fig.~\ref{PaB}). 

One reason why one may not wish to accept the individual ages as stated above at face value is that our modelling of the SF history, in terms of one instantaneous burst, is almost certainly too crude. For instance, Allard et al. (2006) and Sarzi et al. (2007) developed a number of simple multi-burst models which handsomely explain their spectral measurements of the star-forming regions within the circumnuclear rings in several spiral galaxies in terms of a consistent dynamical and evolutionary picture. And in the case of the region in M83 under consideration here, Houghton \& Thatte (2008) find that their observations of individual clusters can be much better reproduced by a model describing a finite episode of SF lasting some 6\,Myr than by an instantaneous one-burst model.

What is surely beyond doubt is the {\it existence of an age gradient} with the youngest stars in the northern, and the oldest in the southern part of the central region. 


\subsection{The {\it SST} near- and mid-IR view}

The {\it SST} near- and mid-IR IRAC images (Fig.~\ref{spitzer}), at wavelengths of 3.6, 4.5, 5.8, and 8.0\,$\mu$m, highlight the complicated interplay between SF and absorption and emission by dust. As far as we are aware, images at the reddest of these wavelengths at the spatial resolution presented here have not been published. The 10\,$\mu$m image presented by Telesco (1988), for example, has a spatial resolution of some 4\,arcsec. The more modern,  extensive, study of Vogler et al. (2005) employed ISOCAM images with wavelengths between 4 and 18\,$\mu$m of which the pixel size was $6\times6$\,arcsec, so that the central region of M83 under scrutiny in the present paper was mostly covered by one pixel. 

Perhaps the most striking feature of especially the bluer of the IRAC images is the absence of the dust lane which is so prominent in the visible and even at 1\,$\mu$m. It is mostly absent from the IRAC images, as expected of course. At 3.6\,$\mu$m and to a lesser extent at 4.5\,$\mu$m the SF on the west side of the central region is seen to extend all the way from region A, in the south, via regions C and D, to region E in the north (Fig.~\ref{spitzer}). The visible nucleus is hardly, if at all, brighter than the arc at these wavelengths, and becomes even less prominent in the redder images. 

SF, through heating of dust and polycyclic aromatic hydrocarbon (PAH) molecules, adds an important element of emission to the redder IRAC images (those at 5.8 and 8.0\,$\mu$m), as is also visible in Telesco's (1988) 10\,$\mu$m image. The youngest region of the star-forming arc, that to the north of the centre, is, in fact, by far the brightest area in the 8.0\,$\mu$m image, which traces predominantly PAH molecules heated by SF (Fig.~\ref{spitzer}). 

This SF-related component of emission is highlighted in the colour index maps which indicate the differences between the maps at the various wavelengths, as shown in Fig.~\ref{spitzer_ratios}. These maps again confirm that the northern region is indeed the one with the most active SF, yet mostly hidden from optical view. This region is red, whereas the southwestern arc which is so prominent at visible wavelengths is relatively inconspicuous, though red because of excess emission at the longest wavelengths in the relevant figure panels. It is slightly puzzling why the arc to the south-southeast of the nucleus should appear so red in all but one of our ratio maps, and it may well indicate another region of SF hidden by significant dust extinction. Our CIRPASS field of view only covers a small part of this arc (see Fig.~\ref{PaB}), but in the extreme lower left corner of the \pab\ map there is a small patch of enhanced emission which might be consistent with SF there. This remains to be confirmed.


\subsection{Kinematics} 

\begin{figure*}
\includegraphics[width=0.9\textwidth]{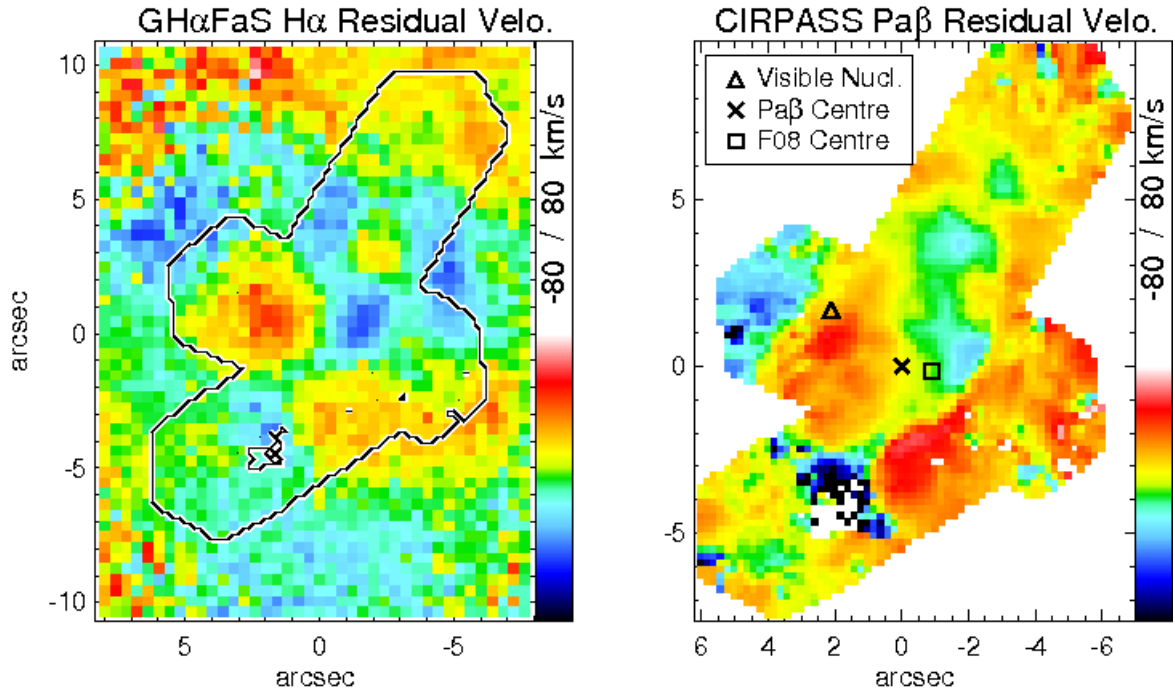}
  \caption{H$\alpha$ residual velocities after removing the
nuclear disk from the GH$\alpha$FaS Fabry Perot interferometer data (left) and the
corresponding non-circular velocities for the Pa$\beta$ emission line (right). The outline of the CIRPASS field is shown in the left panel, and the location of the visible nucleus and of the kinematic centre as derived by Fathi et al. (2008) and from the \pab\ velocity field in the right panel.}
\label{vel_residuals}
\end{figure*}

The ionised gas velocity field derived from our Pa$\beta$
measurements (Fig.~\ref{PaB}) is dominated by rotation, but shows several deviations from circular motion. One way to unveil this is by constructing a model for the velocities in the inner disk using the parameters derived by Fathi et al. (2008), and subtracting it from the \cirpass\ \pab\ velocity
field. We take this approach because the field of view of the CIRPASS data is too small to constrain the overall kinematic parameters, while the \ha\ field of view is very much larger and does allow the construction of a kinematic model which is tightly constrained by the rotation in the inner few kpc area of the disk of M83.

Fathi et al. (2008) report a decoupled nuclear disk located within the inner ILR of the bar, which
rotates significantly faster than
the outer disk, and confirm that the centres of rotation of both the inner and outer disk
coincide to within the uncertainties. The nuclear disk has a scale length of 60\,pc, or
$\sim2.5$\,arcsec, and this is thus the disk of interest for the scales of the
CIRPASS observations. The model is that of an exponential disk, however, the kinematic
parameters (kinematic centre and position angle) were determined by means of harmonic
decomposition of the velocity field (see Fathi et al. 2005 for details). The model confirms
that the nuclear disk dominates the dynamics of the centre of M83. Investigating the
residual velocities in detail, Fathi et al. (2008) found evidence of gas spiralling in from the outer part of this disk, at scales of hundreds of parsecs,
down to the circumnuclear region. They also found that the amplitudes of the non-circular
motions over the extent of the circumnuclear starburst are enhanced. These results are shown in the
third panel of Fig.~\ref{PaB}, which shows the model in the area of the CIRPASS field,
shifted so as to be centred at the position of the cross in the figure to match the slight systemic
velocity offset of a few \kms (see Fig.~\ref{vel_residuals}). Figure~\ref{vel_residuals} shows, in
the left panel, the residual velocities after subtracting the model from the inner part
of the \ha\ velocity field of Fathi et al. (2008) and, in the right panel, our \pab\ velocity field after
subtracting the same model. This confirms that the overall \pab\ velocity field displays dominant disk
rotation, superimposed upon which are important contributions from non-circular motions,
and, more importantly for the discussion here, that the non-circular motions for H$\alpha$ and \pab\ have a similar
behaviour.

The \pab\ kinematic map (Fig.~\ref{PaB}) and the residual field (Fig.~\ref{vel_residuals}) show two prominent regions of high velocity 
gradient, which are located symmetrically to 
the NE and SW of the kinematic and photometric centre (see Sect.~6 for a detailed discussion of the location of the centre).

The most prominent of these, the region to the SW of the kinematic centre, coincides spatially with the SW region of high velocity gradient, the onset of the
prominent star-forming arc to the SW and the dust lane as seen in
the {\it HST} image (Fig.~\ref{cirpass_field}). The gas velocity dispersion is enhanced
in a well-defined arc-shaped ridge to the W of the nucleus, and we
interpret this as a combination of \emph{beam smearing} across the
region of sharp velocity gradients, and of shocked gas along the
leading edge of the dust lane. This is primarily a local feature
related not to the gravitational potential (as inferred by Mast et
al. 2006 and D\'\i az et al. 2006) but to local effects of the dust
lane (see, e.g., Athanassoula 1992 for a detailed discussion of how
shocks occur near dust lanes in bars, and Zurita et\,al. 2004 for an
example of how this can be observed).

Other local deviations from rotation can be seen in the velocity
field, all of minor amplitude, and these can again all be identified
with streaming motions related to the gas flow through the bar and its
dust lanes into the central regions, where SF ensues.

The enhanced stellar velocity dispersion which led Thatte et al. (2000) to the discovery of
the ``second nucleus'' has no significant counterpart in the gas. Although gas and stellar velocity do not necessarily show the same behaviour, our lack of increase near the kinematic centre is in much better agreement with the absence of an increase in the stellar velocity dispersion reported by Houghton \& Thatte (2008). 


\section{Position of the nucleus, or nuclei}

\begin{table*}
\begin{center}
\begin{tabular}{lccl}
\hline
\noalign{\smallskip} 
Method & \multicolumn{2}{c}{Distance from visible nucleus}  & Reference\\
 & $\Delta$(RA) (arcsec) & $\Delta$(dec) (arcsec) &\\
\hline
\noalign{\smallskip}
\pab\ kinematic centre & $-2.1\pm4.8$ & $-1.6\pm4.0$ & This paper\\
G\ha FaS kinematic centre & $-3.0\pm2.0$ & $-1.8\pm2.1$ & Fathi et al. (2008); this paper\\
Photometric centre & $-3.04\pm0.08$ & $-1.88\pm0.09$ & This paper\\
Photometric centre & $-3.05\pm0.30$ & $-1.54\pm0.27$ & Thatte et al. (2000)\\
CO kinematic centre & $-$2.85 & $-$2.4 & Muraoka et al. (2009)\\
\noalign{\smallskip}
\hline
\end{tabular}
\end{center}
   \caption[]{Position of the kinematic and photometric centre of M83 with respect to that of the visible nucleus, as determined through different methods, and by different authors. The reference position, of the visible nucleus, has been determined in the current paper from the continuum peak. All positions are in given in arcsec as offsets in RA and dec. The position of the visible nucleus is given by D\'\i az et al. (2006) as $\alpha=\hmsd 13h37m0.95s, \delta=\dmsd -29d51m55.5s$ (J2000), with a 2$\sigma$ uncertainty of $0.\!\!^{\prime\prime}15$. Muraoka et al. give no error estimate for their CO kinematic centre position.}
\label{centrepos}
\end{table*}

The position and nature of the nucleus of M83 continue to be a matter of considerable debate. As far as we can ascertain, Wolstencroft (1988) was the first to notice explicitly that what we now call the visible nucleus of the galaxy does not coincide with the photometric centre of the galaxy as derived from fitting the near-IR isophotes outside the central region. The offset with the position of the photometric and kinematic centre of M83 has since been confirmed by a variety of authors (as reviewed in detail by Houghton \& Thatte 2008), and, below, we add further confirmation from our data. 

Thatte et al. (2000) reported a possible double nucleus in M83 on the basis of two peaks in stellar velocity dispersion that they found from long-slit data. Other authors have proposed the existence of yet more hidden mass concentrations in the circumnuclear region (Mast et al. 2006; D\'\i az et al. 2006; Rodrigues et al. 2009). But as described by Houghton \& Thatte (2008), their new data led them to the conclusion that no evidence in favour of any hidden mass concentration, including the second nucleus, holds up to further scrutiny.

The implications of the nature and position of the nucleus or nuclei in M83 are wide-ranging. A multiple nucleus would be direct evidence for a recent merger of M83 with a smaller galaxy, an event which could conceivably be at the origin of the enhanced SF activity in the central region. A single but offset nucleus might indicate an $m=1$ perturbation, possibly also related to a recent interactive event. We will explore these issues below on the basis of our new data and a critical review of the literature, and concentrating on the location of the photometric and kinematic centre.   

\subsection{The position of the photometric centre, and dust extinction there}

\begin{figure}
\includegraphics[width=0.5\textwidth]{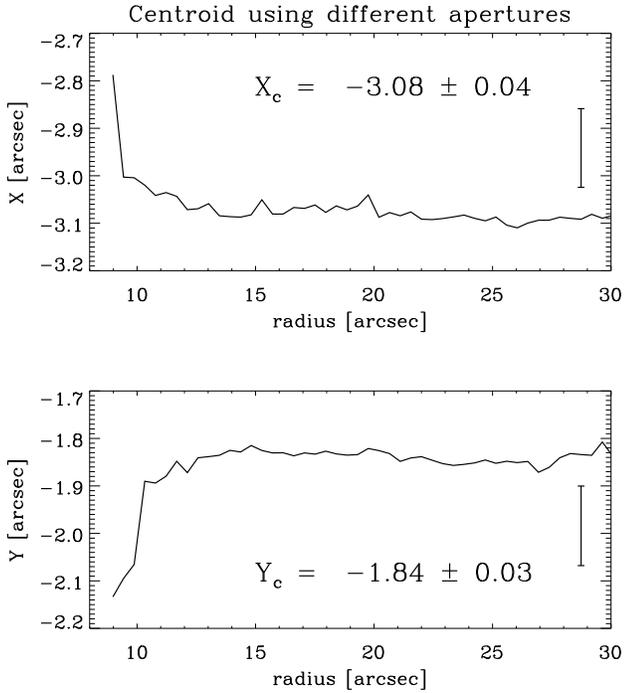}
\caption{Centroid positions as calculated for several ellipses, where the radius of the ellipse is along the abscissa, and the centroid position is given as an offset from the visible nucleus on the ordinate. The mean of the centroids for ellipses between 12 and 30 arcsec in radius is given as $(X_{\rm c}, Y_{\rm c})$.}
\label{centroids}
\end{figure}

First, we will consider the photometric centre of M83, as determined from an extrapolation to the centre of the light distribution at several radii outside the  nuclear region. We will use this information not just to locate the photometric centre, but also to estimate the dust extinction towards it, in the hypothetical case that a standard bulge is hidden from our view by dust. 

To determine the position of the photometric centre we use two different methods, both based on fitting ellipses to the surface brightness of our wide-field AAT $K_{\rm s}$-band image of M83, the central area of which is shown in Fig.~\ref{kmap}. In the first method, we calculate the centroid for several apertures at radii larger than 20\,pixels, or 9\,arcsec. Fig.~\ref{centroids} shows how the fitted centre position changes with radius. The mean value of these coordinates calculated for apertures between 12 and 30\,arcsec then yields the position of the photometric centre, which we find to be $(-3.08\pm0.04, -1.84\pm0.03)$\,arcsec, expressed as an (RA, dec) offset from the position of the visible nucleus. This position is in good agreement with other determinations in the literature, such as those by Wolstencroft (1988), who found an offset of 3\,arcsec at a position angle of 225$^\circ$, and by Thatte et al. (2000) (see Table~\ref{centrepos}). 

\begin{figure}
\includegraphics[width=0.5\textwidth]{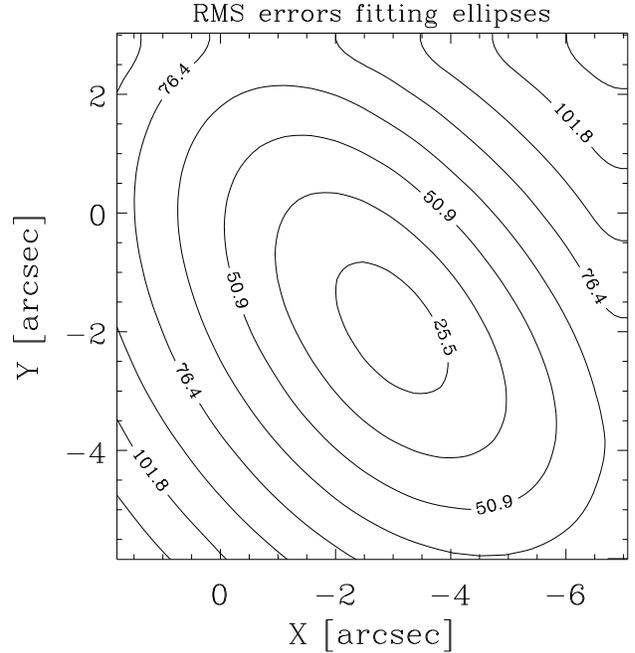}
\caption{Results of a test where a series of ellipse fits was run with the centre position fixed at a grid of $20\times20$ different offsets ($X, Y$). The resulting root-mean-square scatter around the isophotal intensity, in units of counts, is presented here graphically as a function of the centre position.}
\label{rmscontours}
\end{figure}

The second method introduces a variant to the first in the sense that we performed a set of ellipse fits with all parameters left free, with the exception of the position of the centre. The latter was fixed at various values within a grid spanning the central $20\times20$ pixels of the $K_{\rm s}$-band image. This results in 400 individual ellipse fits, with the semi-major axis in each of these ranging from 50 to 100 pixels, far enough out to avoid the nuclear structure, yet far enough in to avoid the noisy outer disk. 

For each ellipse fit, we then calculated the mean of the rms scatter around the isophotal intensity, as given by the parameter `RMS' from iraf's {\sc ellipse} algorithm, considering a number of different isophotes. For each point in the $20\times20$\,pixel matrix we thus find a value for the mean rms, which we show graphically in the form of a contour plot in Fig.~\ref{rmscontours}.  A cubic interpolation of this $20\times20$ matrix into a $100\times100$ matrix then allowed us to fit the optimum position of the photometric centre as $(-2.96\pm0.09, -1.98\pm0.09)$\,arcsec, measured also with respect to the visible nucleus.

By taking the weighted average of the very similar sets of results as derived using these two methods, we arrive at our measure of the position of the photometric centre, which is $-3.04\pm0.08$\,arcsec (west) and $-1.88\pm0.09$\,arcsec (south) of the position of the visible nucleus, where the latter was given by D\'\i az et al. (2006) as $\alpha=\hmsd 13h37m0.95s, \delta=\dmsd -29d51m55.5s$ (J2000), with a 2$\sigma$ uncertainty of $0.\!\!^{\prime\prime}15$. This offset corresponds to one of 79\,pc (in projection, and assuming a distance to M83 of 4.5\,Mpc). This position is the one that we use throughout this paper to denote the position of the photometric centre in the various figures. Our analysis confirms that the position of the visible nucleus of M83 is very significantly offset from that of the photometric centre, or in other words, from the centre of the galaxy as defined by the large-scale photometric structure of the galaxy.

\begin{figure*}
\includegraphics[width=0.45\textwidth]{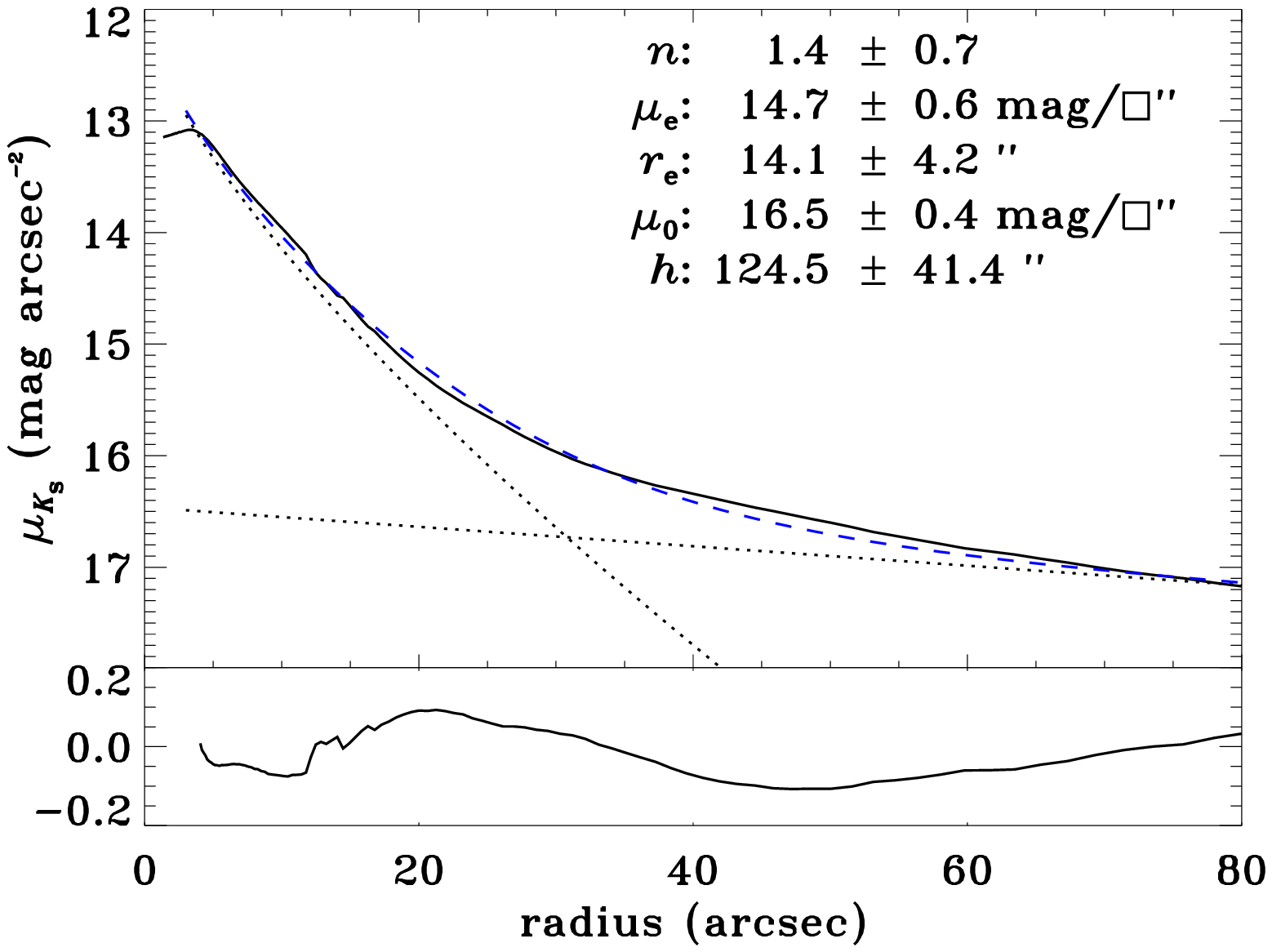}
\includegraphics[width=0.45\textwidth]{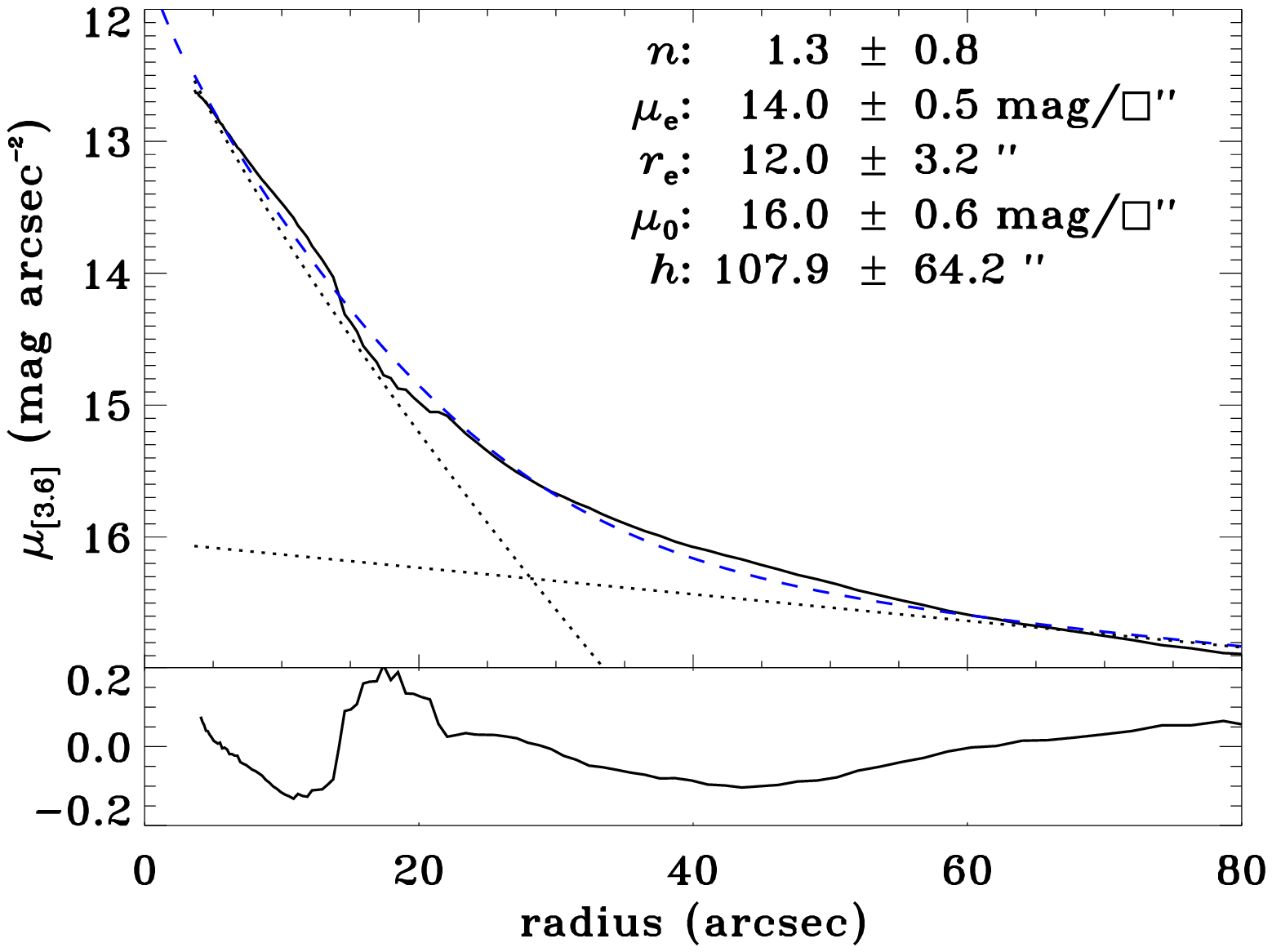}
  \caption{Radial surface brightness profiles as derived from ellipse fitting to
  near-IR images of M83 in the $K_{\rm s}$-band  (left figure) and the 3.6\,$\mu$m {\it SST} band (right figure). The profiles are shown in the top panel (solid curve), along with the 
  bulge and disk components as fitted to it (dotted lines). The blue (dashed) curves in
  the top panels show the combined bulge+disk fits, while the curves in the lower panels 
  are the difference between the radial profile and the fit for each wavelength band. The parameters of the bulge and disk
  fits are given in the upper right corner of each figure.}
\label{SBprofile}
\end{figure*}

The best ellipse fit to the $K_{\rm s}$ light outside a radius of 5\,arcsec results in an exponential disk with a scale length of $125\pm41$\,arcsec and a
Sersic bulge. The Sersic
fit yields $n=1.4\pm0.4$ and $r_{\rm e}=14.1\pm4.2$\,arcsec, and extrapolating it
inwards yields a surface brightness at the position of the true
nucleus which is 1.08\,$K$-mag higher than the value measured in our image (see Fig.~\ref{SBprofile}, left panel).  This implies
an attenuation of 1.08$\pm$0.15\,mag in the $K$-band, or
$A_V\approx9.6$$\pm$1.3\,mag. This amount of extinction, needed to hide the
peak as derived from our Sersic fit to the extent that we observe it
in the $K$-band image, is large but not {\it a priori} unreasonable given the rather
extreme nature of the nuclear region of the galaxy. It is of course possible that the true underlying profile is flattened in the nuclear region, so our estimate of $A_V$ is an upper limit.

\begin{figure}
\includegraphics[width=0.5\textwidth]{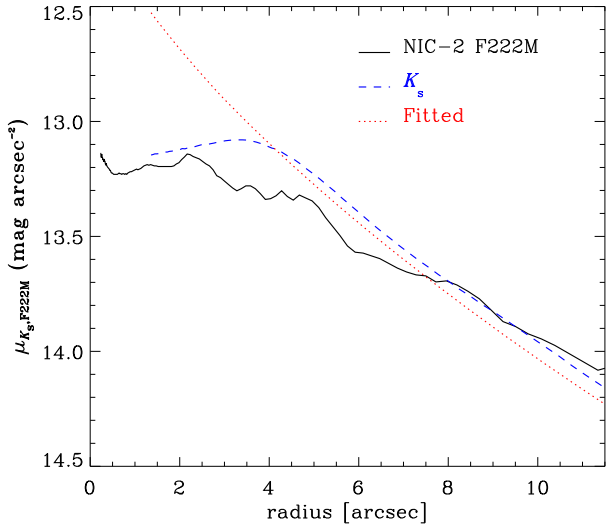}
  \caption{Comparison of the radial surface brightness profiles as derived from our AAT $K_{\rm s}$ (blue dashed curve) and {\it HST} F222M (black curve) images. The extrapolated Sersic fit is shown in red (dotted).}
\label{hstprof}
\end{figure}

\begin{figure}
\includegraphics[width=0.5\textwidth]{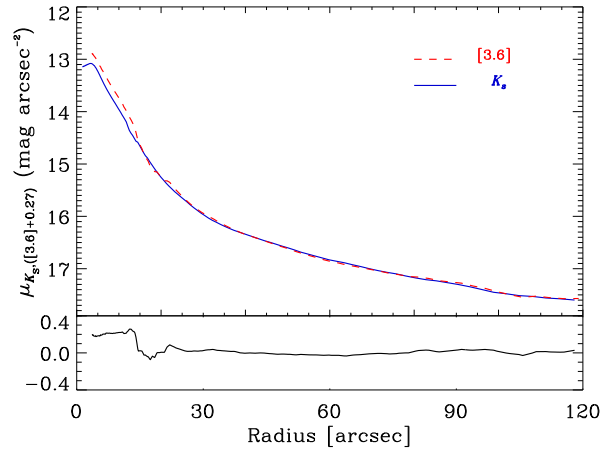}
  \caption{Comparison of the radial surface brightness profiles in $K_{\rm s}$ (blue curve) and at 3.6\,$\mu$m (red, dashed). The 3.6\,$\mu$m profile has been shifted down by $+0.27$\,mag\,arcsec$^{-2}$. The difference between the profiles is shown in the lower panel.}
\label{profilecomp}
\end{figure}

In order to check whether the relatively large pixels and moderate spatial resolution of our AAT $K_{\rm s}$ image might have influenced this result, we made ellipse fits to the surface brightness in the F222M filter in the {\it HST} NICMOS image. This comparison can only be made in the central region because of the limited field of view of the {\it HST} image, and the {\it HST} image is thus not suitable to constrain the overall parameters of the fitted profile. But what Fig.~\ref{hstprof} shows is that even in the very central regions, use of our AAT $K_{\rm s}$-band image is fully satisfactory. In particular, because the surface brightness profile is essentially flat within a radius of some 2\,arcsec, the surface brightness measured from the AAT image at the very photometric centre of the galaxy (of $\sim13.2$\,mag\,arcsec$^{-2}$) is indeed the same value as that derived from the NICMOS image.

Because the pixel size of the {\it SST} near-IR images is over 1\,arcsec we chose to perform the above analysis with our $K_{\rm s}$ image. But the 3.6\,$\mu$m {\it SST} image is very well suited to check the $K_{\rm s}$-band results on the outer disk. We thus fitted ellipses to the 3.6\,$\mu$m image as well, and show the resulting surface brightness in the right panel in Fig.~\ref{SBprofile}. The parameters of the best fit to the bulge and disk component, as shown in the top right corner of the figure, agree well with the values derived from our $K_{\rm s}$-band image. The difference between the $K_{\rm s}$-band and 3.6\,$\mu$m image surface brightness profiles is shown in Fig.~\ref{profilecomp}. There, we see that the difference is very small indeed, $<0.03$\,mag\,arcsec$^{-2}$ over the entire radial range $30-120$\,arcsec. Inside that range in radius, we see a region of excess $K_{\rm s}$ emission between radii of 15 and 22\,arcsec, and one of excess 3.6\,$\mu$m emission inside 15\,arcsec. The former, blue, region would normally be explained by excess SF, but it lies just outside the circumnuclear star-forming region and well inside the bar, which is not a zone which stands out for SF. The latter, red, region can more plausibly be explained by dust extinction obscuring slightly more light at 2.2 than at 3.6\,$\mu$m. 

The main conclusion from this analysis of the 3.6\,$\mu$m image is that the $K_{\rm s}$-band surface brightness profile we used in the above analysis is a reliable tracer of the old stellar population, and not affected by image defects such as incorrect sky subtraction. 

\subsection{The position of the kinematic centre}

\begin{figure}
\begin{center}
\includegraphics[width=0.45\textwidth]{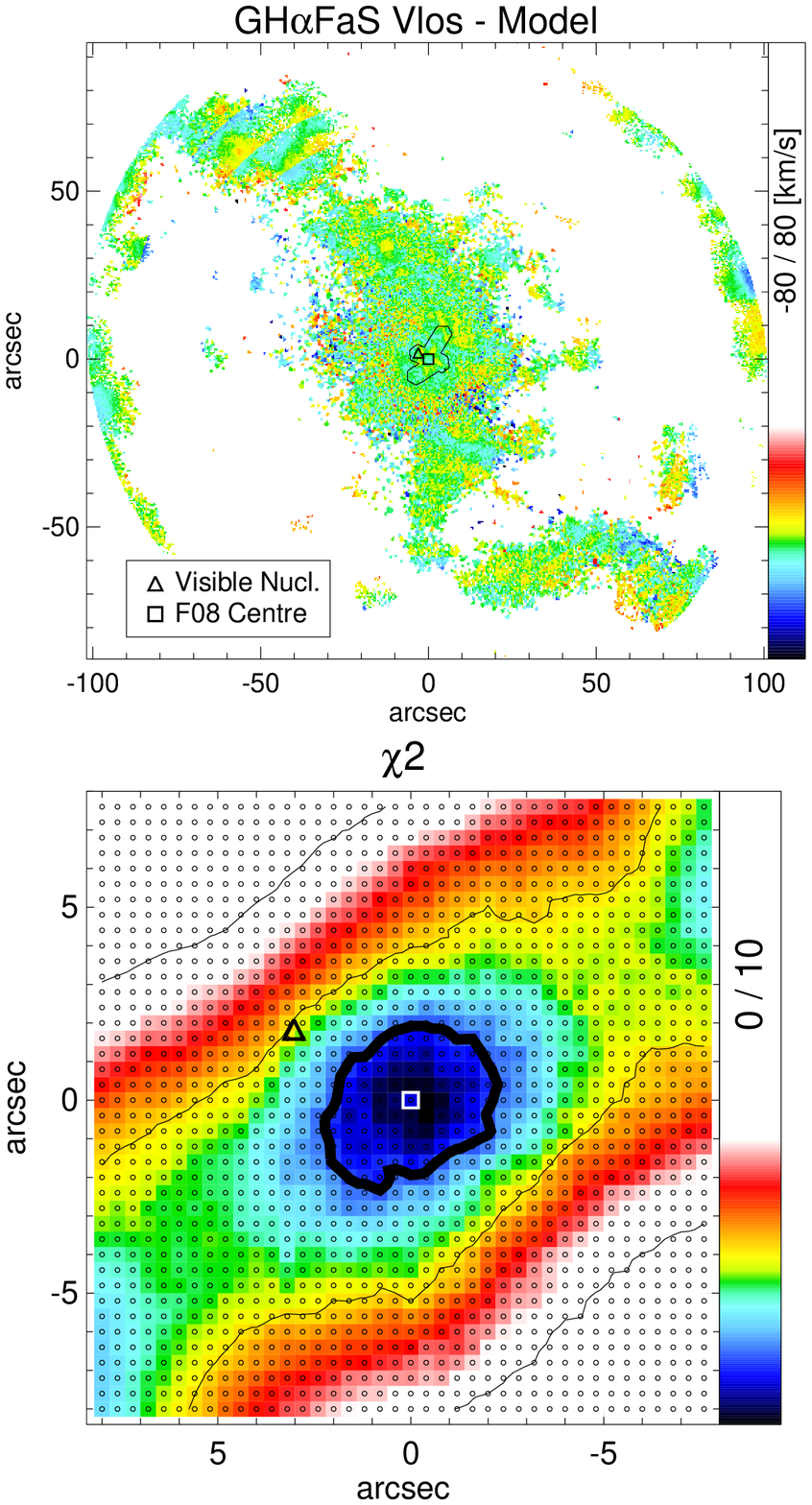}
\end{center}
  \caption{{\it Upper panel}: Residual velocities obtained after subtracting a harmonic decomposition model from the \ha\ velocity field of Fathi et al. (2008). The kinematic centre resulting from the analysis in that paper is indicated by the square, whereas a triangle indicates the position of the visible nucleus. The outline of the area covered by the CIRPASS observations is indicated for reference. {\it Lower panel}: Reduced $\chi^2$ contour map obtained by shifting the harmonic decomposition model of the Fathi et al. (2008) \ha\ velocity field in the RA and dec directions by the amounts (in arcsec) indicated on the $x$- and $y$-axis of the figure. Contour levels are, from inside out, 1 (thick contour), 2, and 3\,$\sigma$. The white square (arbitrarily set at relative position (0,0)) indicates the position of the kinematic centre, the triangle that of the visible nucleus.}
 \label{dyncentre}
\end{figure}

Sakamoto et al. (2004) found that the kinematic centre of M83 ``most likely coincides'' with the position of the photometric centre, and is offset by ``at least'' 3\,arcsec from the visible nucleus. From new Nobeyama Millimeter Array CO observations, Muraoka et al. (2009) confirm this, as shown in our Table~\ref{centrepos}.  We now use our new data (see Sect.~3.4 for a description of the kinematic analysis thereof), and the Fabry-P\'erot \ha\ velocity field obtained with the G\ha FaS instrument by Fathi et al. (2008) which covers a much larger spatial area, to constrain the relative offsets further.

We first show, in the upper panel of Fig.~\ref{dyncentre}, the success of a harmonic
decomposition model in fitting the \ha\ velocity field of Fathi et al. (2008). This model was described
in detail by Fathi (2004) and Fathi et al. (2005). Two of the most notable
features in the current context are the large field of view of the G\ha FaS data,
of some 3\,arcmin diameter, and the success of the model in reproducing the overall
structure in the velocity field. Although the ionized gas is subject to strong winds from
\hii\ regions and other local velocity sources, the overall
galactic-scale velocity field is dominated by the gravitation of the galaxy, and most of
the pixels in the observed velocity field represent the disk potential (see, e.g., Fathi 2004; Falc\'on-Barroso et al. 2006; Fathi et al. 2007; Epinat et al. 2008). Moreover, the
larger-scale effects due to features such as bars and spiral arms are bi-symmetric and hence their
kinematic effect can be represented adequately by bi-symmetric harmonic terms in the harmonic
expansion formalism (see Fathi 2004 for details). This explains why this method can be
successfully used to determine the kinematic parameters, which were derived iteratively for
a fixed inclination and by minimizing the expression $V_\mathrm{los}^2 - (c_1^2 +
s_1^2)$. Similar to the standard tilted-ring method (Begeman 1989), the algorithm minimises the
expression $\chi^2 = V_\mathrm{los}^2 - \sum_{i=1}^{3}(c_i^2 + s_i^2)$, which can be seen as the
idealized case where the line-of-sight velocity $V_{\rm los}$ is the result of a circular cosinusoidal term ($c_1$), a
radial sinusoidal term ($s_1$), and second and third order symmetric terms caused
by possible $m=1$ and $m=2$ gravitational perturbations.

We can now use the harmonic model to constrain the position of the kinematic centre. To
do this, we force the harmonic decomposition model to shift in the $x$ and $y$ direction,
and calculate the reduced $\chi^2$ values (for data$-$model after each step we shift the
model) across a complete grid of offsets spanning $\pm8$\, arcsec in both RA and dec. The
results of this exercise are shown in the lower panel of Fig.~\ref{dyncentre}, and
confirm that the best-fit position of the kinematic centre from the GH$\alpha$FaS field
is significantly offset from that of the visible nucleus, which lies at the 2\,$\sigma$ contour. The offset is
of some 3\,arcsec, or just over 60\,pc, as summarised in
Table~\ref{centrepos}. That table shows also how the location of the photometric centre, as determined in the previous section,
coincides to within the uncertainties with that of the kinematic centre.


\section{Discussion: the nature of the nucleus of M83}

From the data presented in this paper, and considering the results and ideas published in the literature, we can reach two scenarios for the location of the nucleus, or nuclei in M83, which we will discuss critically below. 

\subsection{Stellar velocities}

A significant mass concentration in the central region should lead to a local increase in the stellar velocity dispersion, and this is exactly what has been reported by Thatte et al. (2000). They found evidence for two distinct regions of enhanced velocity dispersion, one coinciding with the visible nucleus, the other very near the location of the photo-center. This has been interpreted by these authors as evidence for a double nucleus in M83, and quoted as such in several papers in the literature. However, no independent confirmation of the increase in velocity dispersion near the photometric centre has been published, and Houghton \& Thatte (2008) have recently shown from new long-slit data that there is in fact no peak in dispersion at this position. The latter authors thus refute the earlier claims, and also claim that there is no longer evidence for the presence of a ``dark'' mass concentration at the photometric centre. 

Other authors (Mast et al. 2006; D\'\i az et al. 2006; Rodrigues et al. 2009) have reported localised areas of rapid change in the line-of-sight velocity in ionized gas velocity fields, and interpreted them as tracers of hidden mass concentrations. We ratify the arguments outlined in detail by Houghton \& Thatte (2008), who do not agree with this interpretation. The effects observed by Mast, D\'\i az, and Rodrigues and their collaborators, over very small fields of view, should be thought of as consequences of local and predominantly non-gravitational effects, including shocks, rather than as small areas of self-consistent rotation tracing the gravitational field of significant local mass concentrations. 

One important missing piece of information is a complete two-dimensional stellar velocity field across the central region. In spite of the abundant observations, such a field has not yet been published. We tried hard to derive it from our current {\sc cirpass} data set using the CO stellar absorption lines, but unfortunately the signal-to-noise ratio is not high enough in most of our pixels to allow us to derive meaningful constraints on the stellar velocity field. Future work should aim to address this critical lack of data. For now, there are two distinct options, which we will discuss below.

\subsection{Option 1: hidden nucleus}

The first option for the location of the nucleus of M83 is that the photometric and kinematic centre do indeed indicate the nucleus. In this case, there is either one (hidden) nucleus, and the visible ``nucleus'' is an extremely massive star cluster forming part of the nuclear ring structure, or there are two, as proposed by Thatte et al. (2000). 

Apart from the rather interesting constraint that it is located at the `centre' of the galaxy (the photometric or isophotal, as well as the kinematic centre), there is, however, rather little support for this model. The original evidence by Thatte et al. (2000) of enhanced velocity dispersion at the position of the photometric and kinematic centre is not reproducible (Houghton \& Thatte 2008) and can be discarded, and attempts to locate an AGN, which could, in principle, be used to confirm the location of the nucleus, remain unsuccessful. Optical or near-IR spectra (including our own) do not show evidence for an AGN, nor does {\it Chandra} X-ray imaging (Soria \& Wu 2003). Radio measures (e.g., those by Maddox et al. 2006) are compromised by the complex environment from which they originate but do not favour the presence of an AGN. 

Circumstantial evidence in favour includes the morphology of the enhanced SF. Considering the visible nucleus to be a very massive star cluster (with a mass of order $10^7\,M_\odot$---comparable to that of the very most massive super star clusters known) and part of the circumnuclear structure, the SF traces a small nuclear ring, roughly centred on the photometric/kinematic centre and with a radius of some 50\,pc---not unreasonable, as such small nuclear rings in other galaxies have been reported in the literature (see, e.g., Comer\'on et al. 2008, 2010). The morphology would be that of a nuclear ring fed by gas that is flowing into the ring under the influence of the bar, a gas flow which is directly illustrated in images by the major dust lanes coming in from the bar. The nuclear ring in M83 would in such a case be very similar to that in, e.g., M100, which can be considered a prototype and was observed and modelled extensively by Knapen et al. (1995a,b, 2000). 

As we have shown above, in Sect.~4.1, it is not implausible {\it a priori} to locate a hidden nucleus at the photometric and kinematic centre. Extrapolating a radial profile fitted to the light distribution outside the central region, we found that dust extinction of $A_V\sim10$\,mag can hide a standard inner part of a Sersic-bulge. In earlier work, however, Gallais et al. (1991) reported (their Fig.~2) that the location of the photometric/kinematic centre in a [$J-H$] vs [$H-K$] diagram is compatible with that of ``normal'' galaxies (following Aaronson 1977), assuming only a moderate amount of extinction by dust, of perhaps 2 or 3 $A_V$ magnitudes. We thus conclude that dust extinction at the position of the photometric/kinematic centre is somewhere in the range $3-10$\,$V$-mag, which may be enough to hide a standard Sersic-type peak in the light profile. This is not unreasonable in principle, given the perturbed nature of the central region, as indicated by the much enhanced massive SF, and the presence of prominent dust lanes. No dust patches can be seen (e.g., in Fig.~\ref{cirpass_field}) at the location of the `hidden nucleus',  however, although the dust might be distributed more uniformly.

The fact that the intermediate age of the visible nucleus (B, see Sect.~3.2) is consistent with it being coeval with the star-forming arc allows us to speculate that the former might be triggered by the same inflow event which feeds the arc, or, indeed, that region B is the only cluster visible on the opposite side of a star-forming ring centred on the photometric and kinematic centre.

We conclude that while we cannot exclude the possibility that there is a hidden nucleus at the position of the photometric and kinematic centre of M83, the lack of enhancement in the optical and near-IR emission and in the stellar velocity dispersion there make this option seem somewhat contrived.

\subsection{Option 2: visible nucleus}

The second option is that the visible nucleus of M83 is indeed its nucleus, but that it is simply offset, by some 80\,pc in projected distance, from the photometric and kinematic centre of the galaxy. Apart from the offset nucleus, the rest of the central region of M83 is relatively normal: the shape of the gas velocity field, the large-scale distribution of stars, the presence of a rather prominent bar which facilitates gas inflow, and a partial  nuclear ring with abundant massive SF. Of these, the velocity field is perhaps most noteworthy as it has been derived from emission from young stars, and thus offers a rather contemporary view of the region. 

Houghton \& Thatte (2008) discuss this model in some detail, and favour it for explaining M83's central region. In particular, they describe how an $m=1$ perturbation can in principle lead to an offset location of the nucleus, and give examples of various other galaxies in the literature with offset nuclei. They also highlight that they could not confirm the enhancement in stellar velocity dispersion reported by Thatte et al. (2000), which the latter authors had interpreted as evidence for the presence of a mass concentration, possibly a second nucleus, at the location of the photo- and kinematic centres. 

One may wonder how a reasonably massive galaxy nucleus (mass of order $10^7\,\msol$, Houghton \& Thatte 2008, even though this mass is only $\sim2\%$ of the mass of the inner rapidly rotating disk of Fathi et al. 2008) can be so offset without apparent major damage to the part of the disk immediately surrounding it (we refer here to the inner, optical/near-IR, disk; the outer disk is rather messy as seen in  \hi---references are given in Sect.~1). In particular, if the offset nucleus is a recent feature of this galaxy, how can even the gas velocity field, derived from emission by stars of only a few million years of age, appear relatively undisturbed by, and insensitive to, the presence of the nucleus? 

A hint at the answer to this question may be provided by the harmonic analysis which Fathi et al. (2008) performed as part of their study of the \ha\ Fabry-Perot gas velocity field of the central  
region of M83. This analysis results in a systematic variation of the zero-th and  
second harmonic terms, which is fully consistent with the presence of an $m=1$  
gravitational perturbation (see Canzian 1993; Swaters et al.  
1999). As ionized H$\alpha$ gas is also subject to considerable local  disturbances in the 
kinematics, for instance in the form of shocks related to non-circular motions or expanding H{\sc ii} regions, we cannot conclude from this analysis that an $m=1$ perturbation is present. The \ha\ dataset of Fathi et al. (2008) is not deep enough (60\,min exposure time) to cover the regions  
outside the bar, so that the variation of the zero-th and/or second  
harmonic terms cannot be traced reliably beyond 20\,arcsec.
The current CIRPASS data set is not useful for this purpose because of its smaller field of view and signal-to-noise ratio. 
A comprehensive analysis of this effect, which may lead to a confirmation of an $m=1$ signature in the velocity field, is only possible using a combination of a stellar velocity field at high spatial and spectral resolution (ideally obtained with adaptive optics techniques), and covering the central  
20-30\,arcsec radius of the disk in M83.

The origin of the offset nucleus, if that is indeed what it is, must be a dramatic external event such as a galaxy-galaxy interaction or a merger. Phenomenologically this is not an unattractive proposition, as such an event could also explain the messy structure seen in \hi\ outside the optical disk (see discussion in the Introduction), and possibly even the much enhanced star-forming activity in the central region. 


\section{Summary and final remarks}

We report new near-IR integral field spectroscopy of the central region of the barred spiral galaxy M83, as obtained with the CIRPASS instrument on the Gemini-South telescope. We analyse the structure and properties of the circumnuclear region from this data set, used in conjunction with an AAT $K_{\rm s}$-band image, near- and mid-IR images obtained from the {\it HST} and {\it SST} archives, and an H$\alpha$ Fabry-Perot velocity field which we have published earlier. Our main conclusions can be summarised as follows.

\begin{itemize}

\item We describe the well-known optical morphology, which is characterised by a prominent large bar with dust lanes outlining the flow of gas from the disk into the central region, where an arc-shaped region of massive SF is seen to the SW of the visible nucleus. From the near-IR \pab\ emission line, we see how the region of massive SF is not limited to the area seen in optical imaging, but instead continues to the north of the nucleus.  

\item Colour index maps based on the {\it SST} IRAC images identify an area to the SSW of the visible nucleus which is characterised by red colours, and which may indicate another region of SF hidden by significant quantities of dust. Since our CIRPASS field of view only skims this area, further observations are needed to confirm this.

\item The bulk of the current SF activity is hidden from optical view by dust extinction, but is seen prominently to the north of the nucleus in the near- and mid-IR. This region is being fed by inflow through the bar of M83, traced by the prominent dust lane entering into the circumnuclear region from the north. From a comparison of our CIRPASS data and stellar population modelling, we confirm that the dust-obscured star-forming regions to the north of the visible nucleus are the youngest, and that a gradual age gradient exists from north to south along the star-forming arc. Additional evidence for the youngest stars occurring to the north is offered by the {\it SST} colours which are reddest to the north, indicating the most recent SF activity.

\item Detailed analyses of the
\pab\ ionised gas kinematics and near-IR imaging confirm that the kinematic centre
coincides with the photometric centre of M83, and that these are offset significantly, by almost 4\,arcsec or 79\,pc in projection, from the visible nucleus of the galaxy. 

\item We discuss two possible options for the location and nature of the nucleus, or nuclei, of M83. The first of these postulates that the kinematic and photometric centre traces a galaxy nucleus hidden by a quantity of dust yielding $A_V=3 -10$\,mag of dust extinction. The lower number in this range is from the literature, while we derive higher values from an extrapolation of a Sersic-type fit to the near-IR surface brightness distribution. While this scenario is in principle feasible, we see it as unlikely that the true nucleus could be located at the 
photometric and kinematic centre of the galaxy without giving away any 
other indication of its presence, such as a peak in velocity dispersion or in 
near-IR or mid-IR emission.

\item We favour a second option, in which the visible nucleus is offset from the kinematic and photometric centre of the galaxy, presumably as a result of some past interaction, possibly related to the event which lies at the origin of the disturbance of the outer disk of the galaxy. We find some indications for a disturbance in the \ha\ velocity field which would confirm the influence of the $m=1$ perturbation in the gravitational potential, but note that further high-quality stellar kinematic data are needed to fully confirm this scenario.

\end{itemize}


\section*{acknowledgments}

We thank Nate Bastian, Judith Croston, and Andreas Lundgren for
stimulating discussions on earlier versions of the manuscript and the anonymous referee for helpful comments. \cirpass\ was built by Ian Parry and the
instrumentation group of the Institute of Astronomy in Cambridge, UK,
co-funded by the Raymond and Beverly Sackler Foundation and
PPARC. KF acknowledges support from the Swedish Research Council (Vetenskapsr\aa det). Based on observations obtained at the Gemini Observatory, which is operated by the
Association of Universities for Research in Astronomy, Inc., under a cooperative agreement
with the NSF on behalf of the Gemini partnership: the National Science Foundation (United
States), the Science and Technology Facilities Council (United Kingdom), the
National Research Council (Canada), CONICYT (Chile), the Australian Research Council
(Australia), Minist\'erio da Ci\^encia e Tecnologia (Brazil) and the Ministerio de
Ciencia, Tecnolog\'\i a e Innovaci\'on Productiva (Argentina). Based on observations made with the AAT and with the
NASA/ESA {\it HST}, the latter obtained from the data archive at the Space
Telescope Science Institute. STScI is operated by the Association of
Universities for Research in Astronomy, Inc. under NASA contract NAS
5-26555. This work is based in part on archival data obtained with the {\it SST}, which is operated by the Jet Propulsion Laboratory, California Institute of Technology under a contract with NASA. 



\bsp

\label{lastpage}

\end{document}